\documentstyle[prd,epsfig,aps,floats]{revtex} 
\draft

\begin{document}
\twocolumn[\hsize\textwidth\columnwidth\hsize\csname
@twocolumnfalse\endcsname
\title{%
\hbox to\hsize{\normalsize\rm November 1997\hfil
Preprint MPI-PTh/97-73}
\vskip 38pt
Photon Dispersion in a Supernova Core}
\author{Alexander Kopf}
\address{Max-Planck-Institut f\"ur Astrophysik,
Karl-Schwarzschild-Str.~1, 85748 Garching, Germany}
\author{Georg Raffelt}
\address{Max-Planck-Institut f\"ur Physik 
(Werner-Heisenberg-Institut),
F\"ohringer Ring 6, 80805 M\"unchen, Germany}
\date{November 18, 1997}
\maketitle
\begin{abstract}
  While the photon forward-scattering amplitude on free magnetic
  dipoles (e.g.\ free neutrons) vanishes, the nucleon magnetic moments
  still contribute significantly to the photon dispersion relation in
  a supernova (SN) core where the nucleon spins are not free due to
  their interaction.  We study the frequency dependence of the
  relevant spin susceptibility in a toy model with only neutrons which
  interact by one-pion exchange.  Our approach amounts to calculating
  the photon absorption rate from the inverse bremsstrahlung process
  $\gamma nn\to nn$, and then deriving the refractive index $n_{\rm
    refr}$ with the help of the Kramers-Kronig relation.  In the
  static limit ($\omega\to 0$) the dispersion relation is governed by
  the Pauli susceptibility $\chi_{\rm Pauli}$ so that $n^2_{\rm
    refr}-1\approx\chi_{\rm Pauli} >0$.  For $\omega$ somewhat above
  the neutron spin-relaxation rate $\Gamma_\sigma$ we find $n^2_{\rm
    refr}-1<0$, and for $\omega\gg \Gamma_\sigma$ the photon
  dispersion relation acquires the form $\omega^2-k^2=m_\gamma^2$. An
  exact expression for the ``transverse photon mass'' $m_\gamma$ is
  given in terms of the $f$-sum of the neutron spin autocorrelation
  function; an estimate is $m_\gamma^2\approx\chi_{\rm Pauli}
  T\Gamma_\sigma$.  The dominant contribution to $n_{\rm refr}$ in a
  SN core remains the electron plasma frequency so that the Cherenkov
  processes $\gamma\nu\leftrightarrow\nu$ remain forbidden for all
  photon frequencies.
\end{abstract}
\vskip2.0pc]


\section{Introduction}

The photon dispersion relation in astrophysical plasmas is usually
dominated by the electromagnetic interaction with the electrons of the
medium.  It was recently claimed~\cite{MS96}, however, that in a
supernova (SN) core the photon interaction with the magnetic moments
of the nucleons yields the dominant contribution to the refractive
index $n_{\rm refr}$.  Because the new contribution has the opposite
sign of the usual plasma term, the photon four-momentum $K$ would
actually become space-like, allowing for the Cherenkov processes
$\gamma\nu\to\nu$ and $\nu\to\nu\gamma$ which could be of great
importance for the neutrino opacities.  Due to a numerical error in
Ref.~\cite{MS96} the overall magnitude of the nucleon magnetic-moment
effect is in fact much smaller~\cite{Raffelt97}, but even after
this correction it is not very much smaller than the electron plasma
effect and thus deserves a closer look.

It is surprising, at first, that the nucleon magnetic moments
contribute at all to the refractive index because the photon
forward-scattering amplitude on fermions with a magnetic moment is
identically zero. Most recently the photon polarization tensor for an
ensemble of noninteracting spin-$\frac{1}{2}$ particles was calculated
in great detail~\cite{ON97} and it was found that indeed the magnetic
moments alone do not produce any contribution to $n_{\rm refr}$.
However, the underlying assumption of a collisionless system is far
from justified in a SN core where the nucleon spin-spin interaction
plays a dominant role. For photon frequencies below the nucleon spin
relaxation rate $\Gamma_\sigma$ the hydrodynamic limit is the
physically appropriate description (not the collisionless limit)
justifying the use of the Pauli susceptibility in Ref.~\cite{MS96}. In
a SN core the spin relaxation rate is likely to be of the same order
as the temperature $T$ \cite{JKRS96}, typical photon frequencies are
also of that order, so that for the entire spectrum of relevant photon
frequencies neither the hydrodynamic nor the collisionless limits are
truly justified. Therefore, an understanding of the photon refractive
index and its frequency dependence requires a more general analysis
than has been offered in either Ref.~\cite{MS96} or~\cite{ON97}.

Perhaps the easiest way to appreciate this point is to consider photon
absorption.  In a collisionless system of neutral fermions
(``neutrons'') with a magnetic moment $\mu_n$ the refractive index is
$n_{\rm refr}=1$ up to order $\mu_n^2$ which implies that there is no
Landau damping, i.e.\ no Cherenkov effect $\gamma n\leftrightarrow n$.
The only photon damping occurs at order $\mu_n^4$ from magnetic
Compton scattering $\gamma n\to n\gamma$. On the other hand, if our
``neutrons'' interact by a spin-dependent force (which for real
neutrons is caused by pion exchange) we have the
inverse-bremsstrahlung absorption process $\gamma n n\to nn$ so that
we do have photon absorption to order $\mu_n^2$. Its rate far
exceeds that of magnetic Compton scattering because in a SN core the
$nn$ interaction rate is large.  By virtue of the Kramers-Kronig
relation one can then derive the associated refractive index $n_{\rm
refr}$ which does not vanish to order $\mu_n^2$.  (Note that we always
take $n_{\rm refr}$ to be a real quantity even though one sometimes
describes absorption by an ``imaginary part of the refractive
index.'')

We proceed in Sec.~\ref{sec:genrelations} with the general photon
dispersion relation in a pure neutron medium in terms of the dynamical
spin-density structure function by virtue of the
fluctuation-dissipation theorem and the Kramers-Kronig relation.  In
Sec.~\ref{sec:semiheur} we use a semi-heuristic expression for the
dynamical spin-density structure function in the long-wavelength limit
to obtain a quantitative estimate of the magnetic-moment refractive
index in a SN core. In Sec.~\ref{sec:summary} we discuss and summarize
our findings.

%
\section{General Relations}
\label{sec:genrelations}
\subsection{Photon Dispersion}
The main idea behind our treatment of the photon dispersion in a
neutron medium is the observation that the photon absorption rate
$\Gamma_{\rm abs}$ is dominated by the inverse-bremsstrahlung process
$\gamma nn \to nn$ which is enabled by the tensor component of the
pion-exchange force between neutrons.  From the absorption rate we can
determine the refractive index $n_{\rm refr}$ with the help of a
Kramers-Kronig relation.  The inverse-bremsstrahlung process itself
can be calculated easily by the usual perturbative methods.  However,
a SN core is so dense and hot that these methods are not obviously
justified~\cite{JKRS96}.  Therefore, it is more useful to begin with a
discussion of photon absorption in the language of linear-response
theory which allows us to identify more general properties of the
photon refractive index than would be apparent on the perturbative
level.  In order to apply our general results to a SN core we will
then, of course, have to take recourse to a semi-heuristic
perturbative approach to $nn$ interactions (Sec.~\ref{sec:semiheur}).

Photon dispersion is caused by the medium's response to applied
electromagnetic fields. In the homogeneous and stationary case all
relevant information is contained in the polarization tensor
$\Pi^{\mu\nu}(K)$ where $K=(\omega,{\bf k})$ is the frequency and
wavevector of the applied electromagnetic perturbation.  If the medium
is isotropic and parity conserving the polarization tensor is uniquely
characterized by a pair of two response functions which are often
chosen to be the longitudinal and transverse polarization functions
$\pi_L=(1-\omega^2/k^2)\Pi^{00}$ and $\pi_T=\frac{1}{2}({\rm
  Tr}\,\Pi-\pi_L)$ with $k=|{\bf k}|$ the wavenumber of the
perturbation. Because of the assumed isotropy all quantities depend
only on the wavenumber $k$, not on the direction of ${\bf k}$. The
dispersion relation of propagating modes is determined by
$\omega^2-k^2=\pi_{T,L}(\omega,k)$.

Another pair of medium response function is the dielectric
permittivity $\epsilon$ and the magnetic permeability $\mu$ which give
us the displacement ${\bf D}=\epsilon{\bf E}$ caused by an applied
electric field and the magnetic field ${\bf H}=\mu^{-1}{\bf B}$ caused
by an applied magnetic induction. However, the magnetic field ${\bf
H}$ and the transverse part of the displacement, characterized by
${\bf k}\cdot {\bf D}_T=0$, do not have independent
meaning~\cite{Kirzhnits87}. Therefore, among other possibilities one
may equally well choose ${\bf H}={\bf B}$, ${\bf D}_T=\epsilon_T{\bf
E}_T$, and ${\bf D}_L=\epsilon_L{\bf E}_L$ with $\epsilon_L\equiv
\epsilon$ the longitudinal and
$\epsilon_T\equiv\epsilon+(1-\mu^{-1})\,k^2/\omega^2$ the transverse
dielectric permittivity.  They are related to the polarization
functions by $\epsilon_L=1-\pi_L/(\omega^2-k^2)$ and
$\epsilon_T=1-\pi_T/\omega^2$~\cite{Weldon82}.  In this language the
dispersion relations take on their standard form
$\epsilon_L(\omega,k)=0$ and
$\omega^2\epsilon_T(\omega,k)=k^2$~\cite{Sitenko67}.

We are presently only concerned with the dispersion relation of
transverse modes (``photons'') because a medium consisting of magnetic
dipoles is not expected to support longitudinal modes (longitudinal
plasmons). From the above it follows immediately that the photon
dispersion relation can be written in the form
\begin{equation}\label{eq:classicaldisp}
\frac{k^2}{\omega^2}=\epsilon(\omega,k)\,\mu(\omega,k).
\end{equation}
With the usual definition of the photon refractive index
\begin{equation}\label{eq:defn}
n_{\rm refr} \equiv \frac{k}{\omega}
\end{equation}
we arrive at the classical result $n^2_{\rm
refr}=\epsilon\mu$~\cite{Jackson}.  It follows that the refractive
index must be determined self-consistently as a solution of
\begin{equation}
n_{\rm refr}^2(\omega,k)=\epsilon(\omega,k)\,\mu(\omega,k)
\end{equation}
with $k=n_{\rm refr}\omega$ for any frequency $\omega$ of a
propagating mode.  
Depending on the properties of the medium the
long-wavelength approximation
$\epsilon(\omega,k)\approx\epsilon(\omega,0)$ and
$\mu(\omega,k)\approx\mu(\omega,0)$ may be justified, leading to the
much simpler dispersion relation $n_{\rm
refr}^2(\omega)=\epsilon(\omega,0)\,\mu(\omega,0)$.

Sometimes it will be more useful to write the photon dispersion
relation in the form $\omega^2-k^2=m_{\rm eff}^2$ in terms of a
frequency-dependent ``effective mass'' 
\begin{equation}\label{eq:effmass}
m_{\rm eff}^2(\omega)=(1-n_{\rm refr}^2)\,\omega^2,
\end{equation}
where in fact $m_{\rm eff}^2<0$ is possible. For electric interactions
and frequencies well above all resonances we obtain the well-known
plasma effect dispersion relation which implies that $m_{\rm eff}^2>0$
and independent of frequency~\cite{Jackson}.  We will show that the
same holds true for our magnetic case.  Therefore, it is useful to
define
\begin{equation}
m_\gamma\equiv \lim_{\omega\to\infty} m_{\rm eff}(\omega)
\end{equation}
as the (transverse) ``photon mass'' in the medium.

We will mostly be concerned with a medium of neutrons which interact
with the electromagnetic field by their magnetic dipole moment.  In
the nonrelativistic limit they do not respond at all to an applied
electric field so that we may use the approximation $\epsilon=1$.  The
magnetic permeability can be written as $\mu=1+\chi$ in terms of the
magnetic susceptibility $\chi$.  (We use rationalized units where
$\alpha=e^2/4\pi \approx 1/137$ or else we would have to write
$\mu=1+4\pi\chi$~\cite{Jackson}.)

In general the magnetic susceptibility is a complex function of the
real variables $\omega$ and $k$.  Following common practice we write
it in the form
\begin{equation}
\chi(\omega,k) = \chi'(\omega,k) + i\,\chi''(\omega,k)
\end{equation}
in terms of its real and imaginary parts. It is found that that
$\chi''$ is an odd function of $\omega$ while $\chi'$ is
even~\cite{Forster}.  Because we have defined the refractive index to
be a real quantity the dispersion relation is
\begin{equation}\label{eq:nchi}
n_{\rm refr}^2(\omega,k) - 1 = \chi'(\omega,k)
\end{equation}
with $k=n_{\rm refr}\omega$. 

The imaginary part of the susceptibility describes dissipation:
Usually one pictures a stationary beam of frequency $\omega$ along the
$z$-direction which is characterized by a (real) wavenumber $k=n_{\rm
  refr}\omega$ and a damping wavenumber $\kappa=\frac{1}{2}
\lambda^{-1}$ with $\lambda$ the photon mean free path.  The amplitude
of this beam varies as $e^{-i(\omega t - k z)-\kappa z}$, its
intensity as $e^{-2\kappa z}=e^{-z/\lambda}$.  The relativistic limit
$|n_{\rm refr}-1| \ll 1$ implies $n_{\rm refr}^2-1=(n_{\rm
  refr}+1)(n_{\rm refr}-1) \approx 2(n_{\rm refr}-1)$ or $n_{\rm
  refr}-1\approx \frac{1}{2}\chi$.  Therefore, one can picture
$\frac{1}{2}\chi''$ to be an ``imaginary part of the refractive
index'', leading to the identification $\kappa=
\frac{1}{2}\chi''\omega$ or $\chi''=(\lambda\omega)^{-1}$.

We stress that at finite temperature this simple interpretation is not
complete because the medium can both absorb and spontaneously 
emit photons. The two processes are related by the usual 
detailed-balance factor $e^{-\omega/T}$. What is actually damped is
not a mode $k$ of the electromagnetic field, but rather the deviation
of its occupation number from a thermal distribution. It is easy to
show that this damping occurs at a rate $1-e^{-\omega/T}$ times the 
``naive'' absorption rate $\Gamma_{\rm abs}$~\cite{Weldon83}. 
Therefore, the appropriate interpretation of the imaginary part of the
susceptibility is 
\begin{equation}
\chi''(\omega,k) = \frac{1}{\omega}\left(1-e^{-\omega/T}\right)\,
\Gamma_{\rm abs}(\omega)
\end{equation}
with $k=n_{\rm refr}\omega$.  In the limit $|n_{\rm refr}-1|\ll 1$ the
``naive'' absorption rate is $\Gamma_{\rm abs}=\lambda^{-1}$; it is
given by the standard formula ``absorption cross section times target
density.'' 

\subsection{Fluctuation-Dissipation Theorem}

To lowest order the neutrons can absorb photons only because they
interact by a spin dependent force which enables the
inverse-bremsstrahlung process $\gamma nn\to nn$. At the same time
this spin-dependent force causes the neutron spins to fluctuate. The
relationship between spin fluctuations and the absorptive part of the
spin susceptibility is encapsuled in the fluctuation-dissipation
theorem which will help us to understand some general properties of
the photon refractive index.

In our case the most useful quantity to describe the neutron spin
fluctuations is the dynamical spin-density structure function.
Following the normalization convention of Ref.~\cite{JKRS96} it is
defined as
\begin{equation}
S_{\sigma}(\omega,{\bf k}) = \frac{4}{3 n_n}\int_{-\infty}^{+\infty}
dt\, e^{i\omega t}\langle
\bbox{\sigma}(t,{\bf k})\cdot\bbox{\sigma}(0,-{\bf k}) \rangle,
\label{eq:defofS}
\end{equation}
where $\bbox{\sigma}(t,{\bf k})$ is the spatial Fourier transform of
the neutron spin-density operator $\bbox{\sigma}(t,{\bf r}) =
\frac{1}{2} \psi^{\dagger}(t,{\bf r}) \bbox{\tau} \psi(t,{\bf r})$.
Here $\psi(t,{\bf r})$ is a Pauli two-spinor describing the nucleon
field and $\bbox{\tau}$ is a vector of Pauli matrices. Further, $n_n$
is the neutron number density and $\langle\ldots\rangle$ denotes a
thermal ensemble average. Of course, in an isotropic system the
structure function depends only on $k=|{\bf k}|$.  The normalization
was chosen such that
\begin{equation}\label{eq:Snorm}
\int_{-\infty}^{+\infty} \frac{d\omega}{2\pi}\,S_{\sigma}(\omega,0)
= 1
\end{equation}
for a case where there are no static spin-spin correlations between
different neutrons which are taken to be nondegenerate.  In the limit
of vanishing spin-spin interactions we have
\begin{equation}\label{eq:Sdelta}
S_{\sigma}(\omega,0)\to2\pi\delta(\omega).
\end{equation}
Moreover, it satisfies
\begin{equation}
S_{\sigma}(-\omega,-{\bf k}) =
e^{-\omega/T}S_{\sigma}(\omega,{\bf k})
\label{eq:Sdetbalcond}
\end{equation}
as required by the principle of detailed balance.

We next observe that the operator for the magnetization density for
neutrons is ${\bf M}=2\mu_n\bbox{\sigma}$ where the factor 2 is the
gyromagnetic ratio for a spin-$\frac{1}{2}$ particle and $\mu_n$ is
the neutron magnetic moment, not to be confused with the magnetic
permeability $\mu$ of the previous section. A relationship between the
dissipative part of $\chi$ and spontaneous spin fluctuations can then
be written in the form~\cite{Forster}
\begin{equation}
\chi''(\omega,{\bf k})=
\frac{1}{2}\int_{-\infty}^{+\infty}
dt\, e^{i\omega t}\biggl\langle
\frac{1}{3}\sum_{i=1}^3\bigl[M_i(t,{\bf k}),M_i(0,-{\bf k})\bigr]
\biggr\rangle,
\end{equation}
where $[\,\cdot\,{,}\,\cdot\,]$ is the usual commutator. Comparing
this with Eqs.~(\ref{eq:defofS}) and (\ref{eq:Sdetbalcond}) reveals
that this relationship is equivalent to
\begin{equation}
\chi''(\omega,{\bf k}) = \frac{1}{2}\mu_n^2 n_n 
\left(1 - e^{-\omega/T}\right)\,S_{\sigma}(\omega,{\bf k}).
\label{eq:flucdiss}
\end{equation}
In this form it is known as the fluctuation-dissipation
theorem~\cite{Forster,Pines}.


\eject

\subsection{Kramers-Kronig Relation}

Once the imaginary part of the magnetic susceptibility is known the
real part can be found by virtue of the well-known Kramers-Kronig
relation
\begin{equation}
\chi'(\omega,k) = {\rm P}\!\int_{-\infty}^{+\infty}
\frac{d\tilde\omega}{\pi}\, 
\frac{\chi''(\tilde\omega,k)}{\tilde\omega - \omega}
\label{eq:KramersKronigchi}
\end{equation}
where ${\rm P}$ denotes a Cauchy principal value integral.  With the
help of the fluctuation-dissipation theorem Eq.~(\ref{eq:flucdiss}) we
find a direct relationship between the dispersive part of the magnetic
susceptibility and the spin-density structure function
\begin{equation}
\chi'(\omega,k) = 2 \mu_n^2 n_n\, {\rm P}\!
\int_{-\infty}^{+\infty}
\frac{d\tilde\omega}{2\pi}\, 
\frac{\tilde\omega\,S_\sigma(\tilde\omega,k)}
{{\tilde\omega}^2 - \omega^2}.
\label{eq:chiSigma}
\end{equation}
One may use detailed balance to write this in the form
\begin{equation}
\chi'(\omega,k)=\chi_{\rm Pauli}\,{\rm P}\!
\int_{0}^{\infty}
\frac{d\tilde\omega}{\pi}\, 
\frac{1-e^{\tilde\omega/T}}{\tilde\omega/T}\,
\frac{{\tilde\omega}^2\,S_\sigma(\tilde\omega,k)}
{{\tilde\omega}^2 - \omega^2},
\label{eq:chiSigmaB}
\end{equation}
where
\begin{equation}
\chi_{\rm Pauli}\equiv\frac{\mu_n^2 n_n}{T}
\end{equation}
is the usual Pauli susceptibility for a system of collisionless
spin-$\frac{1}{2}$ particles with a magnetic moment $\mu_n$. 


\subsection{Limiting Cases}
\label{sec:limitingcases}

In order to understand the general behavior of the refractive index we
begin with the static limit $\omega \rightarrow 0$. The static
susceptibility $\chi_0(k)\equiv \chi'(0,k)$ has no imaginary part
because $\chi''$ is an odd function of $\omega$. From
Eq.~(\ref{eq:chiSigmaB}) we find for the real part
\begin{equation}\label{eq:staticsus}
\chi_0(k) = \chi_{\rm Pauli}\,
\int_{0}^{\infty}
\frac{d\tilde\omega}{\pi}\, 
\frac{1-e^{-\tilde\omega/T}}{\tilde\omega/T}\,
S_{\sigma}(\tilde\omega,k).
\end{equation}
In the collisionless limit the structure function becomes narrowly
peaked around $\omega=0$. With the help of Eq.~(\ref{eq:Sdelta}) we
thus recover the usual long-wavelength result result
$\chi_0(0)=\chi_{\rm Pauli}$. When $S_\sigma(\omega,k)$ is not
narrowly peaked on scales of the temperature, the static
susceptibility decreases relative to the Pauli value---we show this
effect explicitly in Fig.~\ref{fig:static} below in the framework of a
heuristic toy model.

How large may the frequencies be that the static result is still
approximately justified? The structure function in the long-wavelength
limit $S_\sigma(\omega,0)$ has the interpretation of the
autocorrelation function of a single neutron spin. Therefore, it is a
broad, decreasing function of $\omega$ with a width representing
something like the spin-relaxation or spin-fluctuation rate
$\Gamma_\sigma$. If the external electromagnetic perturbation has a
frequency much less than this, $\omega\ll\Gamma_\sigma$, we are in the
hydrodynamic limit where the neutron spins may fully relax to a new
thermodynamic equilibrium state on the time scale of a period of the
perturbation. In this case we may use the static susceptibility to
estimate the photon refractive index. Moreover, even though we just
saw that the static susceptibility is not independent of the width of
$S_\sigma(\omega)$, this dependence is weak so that in the
hydrodynamic limit the Pauli susceptibility is a good estimate,
justifying the approach of Ref.~\cite{MS96} to photon dispersion in
the limit $\omega\ll\Gamma_\sigma$.

The opposite limiting case is that of very large $\omega$. If
$S_\sigma(\omega)$ falls off sufficiently fast beyond some frequency
$\omega_0$ which is determined by the nature of the $nn$ interaction
potential, then for $\omega\gg\omega_0$ the integral in
Eq.~(\ref{eq:chiSigma}) is dominated by $|\tilde\omega|\alt\omega_0$,
leading to
\begin{equation}
\chi'(\omega,k)
=-2\,\frac{\mu_n^2 n_n}{\omega^2}
\int_{-\infty}^{+\infty} \frac{d\tilde\omega}{2\pi}\,
\tilde\omega\,S_\sigma(\tilde\omega,k).\label{eq:fsumintro}
\end{equation}   
The integral in this equation is the so-called $f$-sum of the
structure function. Independently of the nature of the assumed $nn$
interaction the $f$-sum always exists and is given as a thermal
expectation value of the tensor part of the $nn$ interaction
potential~\cite{Sigl}. Moreover, the $f$-sum is always positive
because of the detailed-balance property Eq.~(\ref{eq:Sdetbalcond}).

For photon dispersion, this result corresponds to a positive value for
the squared effective mass defined in Eq.~(\ref{eq:effmass}). With 
Eq.~(\ref{eq:nchi}) we find
\begin{equation}\label{eq:photonmass}
m_{\rm eff}^2 = 2 \mu_n^2 n_n
\int_{-\infty}^{+\infty} \frac{d\tilde\omega}{2\pi}\,
\tilde\omega\,S_\sigma(\tilde\omega,k).
\end{equation}
If the momentum dependence of this expression is weak so that we may
use the long-wavelength limit then the photon dispersion relation is
that of a massive particle $\omega^2-k^2=m_\gamma^2$ with the
transverse photon mass given by Eq.~(\ref{eq:photonmass}) with $k=0$
on the right-hand side.  The appearance of this form has the same
cause as in the usual plasma case, i.e.\ $n_{\rm refr}$ is given by
the $f$-sum of the relevant dynamical structure functions.

The Pauli susceptibility is a positive number (the neutrons are a
paramagnetic medium) so that in the hydrodynamic limit the photon
dispersion relation is approximately characterized by $n_{\rm
  refr}^2-1=\chi_{\rm Pauli}$ or $m_{\rm eff}^2=-\chi_{\rm
  Pauli}\omega^2<0$. On the other hand in the large-frequency limit we
have $m_{\rm eff}^2>0$ as given in Eq.~(\ref{eq:photonmass}).
Moreover, on dimensional grounds the $f$-sum must take on the
approximate value $\Gamma_\sigma$. Therefore,
\begin{equation}
m_{\rm eff}^2\approx\chi_{\rm Pauli}
\times\cases{-\,\omega^2&for $\omega\ll\Gamma_\sigma$\cr
+\,T\Gamma_\sigma&for $\omega\gg\Gamma_\sigma$\cr}
\end{equation}
gives us a rough picture of the behavior of the photon dispersion
relation in a medium of neutron spins.

\eject


\section{Semi-Heuristic Model}
\label{sec:semiheur}

In a SN core neither the collisionless nor the hydrodynamic limits are
appropriate so that we need to come up with a concrete expression for
the dynamical spin-density structure function in order to estimate the
photon refractive index. In a dilute medium one may use the usual
perturbative methods to compute the processes $\gamma
nn\leftrightarrow nn$. Because the relevant photon energies are small
compared with the neutron mass the momentum transfer of the radiation
to the neutron system may be neglected, an approximation which amounts
to the long-wavelength limit which we shall henceforth adopt with the
notation $S_\sigma(\omega)\equiv\lim_{k\to 0} S_\sigma(\omega,k)$.
Next, one may extract $S^{(1)}_\sigma(\omega)$, where the superscript
indicates that this is a lowest-order perturbative
result. Independently of the details of the assumed $nn$ interaction
potential one finds the generic representation~\cite{JKRS96}
\begin{equation}
S^{(1)}_\sigma(\omega)=\frac{\Gamma_\sigma}{\omega^2}\,s(\omega/T)
\end{equation}
where $s(x)$ with $x=\omega/T$ is a slowly varying function of order
unity. This factorization is somewhat arbitrary; we define what we
call the neutron ``spin-fluctuation rate'' $\Gamma_\sigma$ such that
for nondegenerate neutrons $s(0)=1$. Moreover, we have
\begin{equation}\label{eq:sdetbal}
s(-x)=s(x)\,e^{-x}
\end{equation}
so that the detailed-balance relation for $S_{\sigma}(\omega)$ 
Eq.~(\ref{eq:Sdetbalcond}) is satisfied. 

The lowest-order perturbative representation $S^{(1)}_\sigma(\omega)$
diverges at $\omega=0$ and thus violates the normalization rule
Eq.~(\ref{eq:Snorm}).  However, including multiple-scattering effects
suggests the ``resummed'' representation~\cite{JKRS96}
\begin{equation}\label{eq:resummedS}
S_\sigma(\omega)=\frac{\Gamma_\sigma}{\omega^2+\Gamma^2/4}
\,s(\omega/T).
\end{equation}
In a very dilute medium this function is strongly peaked around
$\omega=0$ so that it approaches $2\pi\delta(\omega)$. In this limit
we have $\Gamma=\Gamma_\sigma$, i.e.\ we approach the classical limit
of a Lorentzian correlation function
$S_\sigma(\omega)=\Gamma_\sigma/(\omega^2+\Gamma_\sigma^2/4)$.  We
stress that the representation Eq.~(\ref{eq:resummedS}) is completely
general if we interpret $\Gamma$ as a function of $\omega$ which in
linear-response theory is related to the neutron spin's ``memory
function'' \cite{Forster}.  In our heuristic description, however, we
will use a constant value for $\Gamma$ which is fixed by the
normalization requirement Eq.~(\ref{eq:Snorm}).

In order to calculate $\Gamma_\sigma$ and $s(x)$ in a dilute neutron
medium we model the $nn$ interaction by one-pion exchange in Born
approximation, an approach which has been common practice for SN and
neutron-star physics since Friman and Maxwell's seminal
paper~\cite{FM79} and which is further justified in Ref.~\cite{HR97}.
Further, we take the neutrons to be nondegenerate which is not a bad
approximation during the early phases of SN core cooling.  Finally, we
neglect the mass in the pion propagator which is also a reasonable
approximation for the large momentum transfers in typical $nn$
collisions in a SN core. All of these approximations go in the same
direction of somewhat overestimating the $nn$ spin interaction rate.
We also ignore static spin-spin correlations which could, in
principle, both enhance or diminish our results.

\begin{figure}[b]
\hbox to\hsize{\hfil\epsfig{file=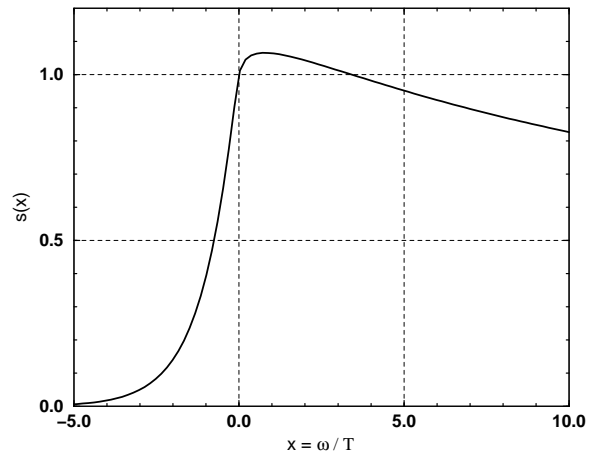,width=6.0cm,angle=270}\hfil}
\smallskip
\caption{\label{fig:sx} The function $s(x)$ as defined in 
Eq.~(\ref{eq:sx}). The analytic approximation Eq.~(\ref{eq:sapp}) is
identical to within plotting accuracy.}
\end{figure}

Within this framework the spin-fluctuation rate is explicitly
found to be~\cite{HR97,RS95}
\begin{equation}\label{eq:gamsig}
\Gamma_\sigma=4\sqrt{\pi}\,\alpha_\pi^2 n_n T^{1/2} m_N^{-5/2},
\end{equation}
where $\alpha_\pi\equiv (f2m_N/m_\pi)^2/4\pi \approx15$ with $f\approx
1$ is the pion-nucleon ``fine-structure constant,'' $n_n$ is the
neutron density, and $m_N$ the nucleon mass. Numerically
we find
\begin{equation}\label{eq:numericalgamma}
\gamma_\sigma\equiv\Gamma_\sigma/T
=8.6\,\rho_{14}\,T_{10}^{-1/2},
\end{equation}
where $\rho_{14}\equiv\rho/10^{14}\,{\rm g\,cm^{-3}}$ and
$T_{10}\equiv T/10\,{\rm MeV}$. Moreover, one finds~\cite{HR97,RS95}
\begin{eqnarray}\label{eq:sx}
s(x)&=&\int_{\max(0,-x)}^\infty dv\,\,e^{-v}
\biggl[\sqrt{v(v+x)}\nonumber\\
&&\hskip3em 
-\,\frac{x^2}{2(2v+x)}\,
\log\left(\frac{\sqrt{v+x}+\sqrt{v}}{\sqrt{v+x}-\sqrt{v}}\right)
\biggr],
\end{eqnarray}
an expression which indeed fulfills the detailed balance requirement
Eq.~(\ref{eq:sdetbal}) and which is smooth at $x=0$ with the
derivative $s'(0)=1/2$ (Fig.~\ref{fig:sx}).  We will use a simple
analytic approximation to this integral~\cite{HR97}
\begin{equation}
s(x)\Big|_{x\geq0}
\approx \left(\frac{x}{4\pi}+\left[1+\left(12+\frac{3}{\pi}\right)x
\right]^{-1/12}\right)^{-1/2}\label{eq:sapp}
\end{equation}
which reproduces the correct limiting behavior for $x\gg 1$ and for
$x=0$ where it also has the correct derivative. It deviates from the
true value by no more than 2.5\% anywhere. For $x<0$ we use
$s(x) = s(-x)\,e^{x}$ in accordance with detailed balance.

\begin{figure}
\vbox{
\hbox to\hsize{\hfil\epsfig{file=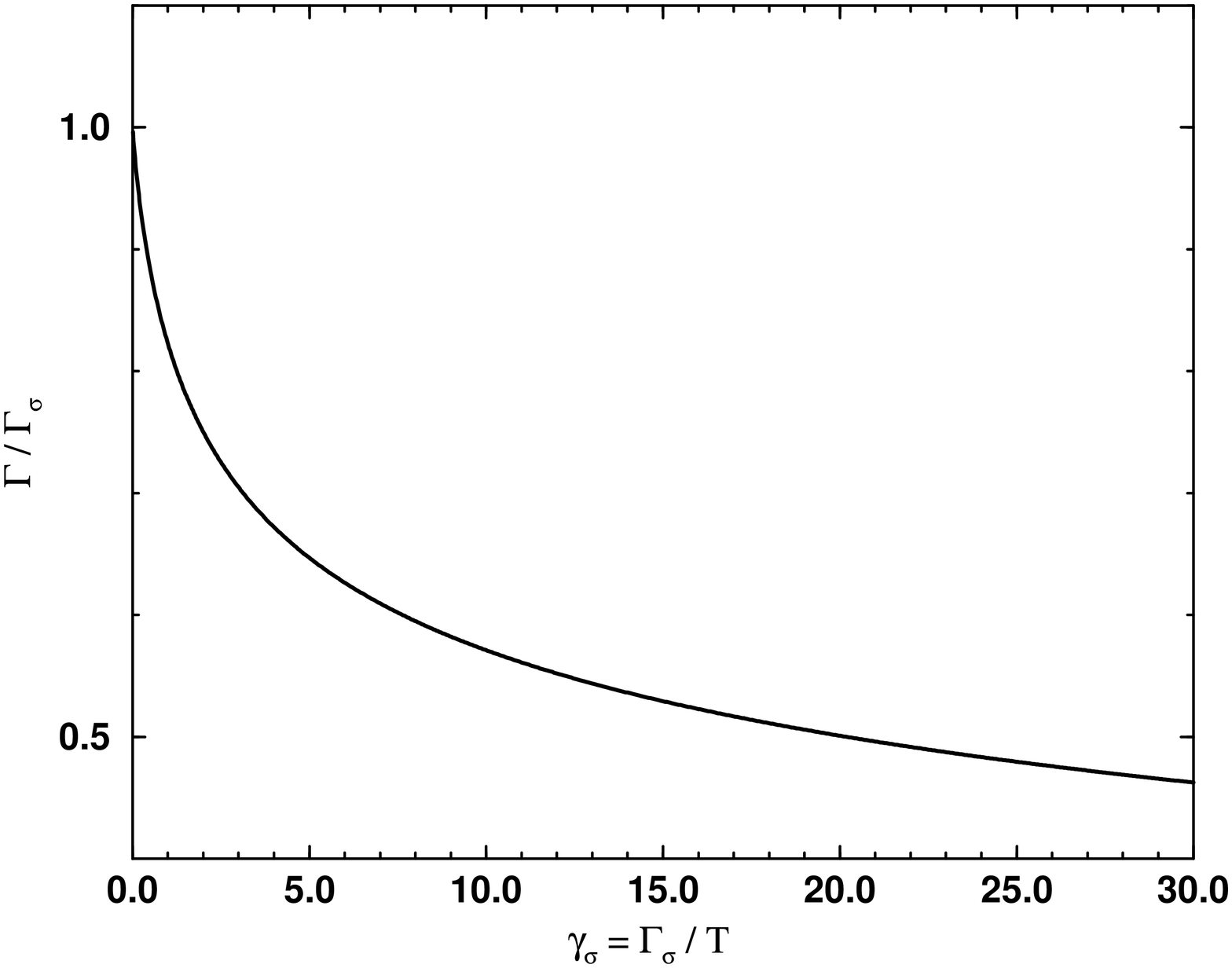,width=6.0cm,angle=270}\hfil}
\smallskip
\caption{\label{fig:norm}  $\Gamma$ appearing in 
Eq.~(\ref{eq:resummedS}) to normalize $S_\sigma(\omega)$ 
as a function of $\Gamma_\sigma$.}
 \bigskip
\hbox to\hsize{\hfil\epsfig{file=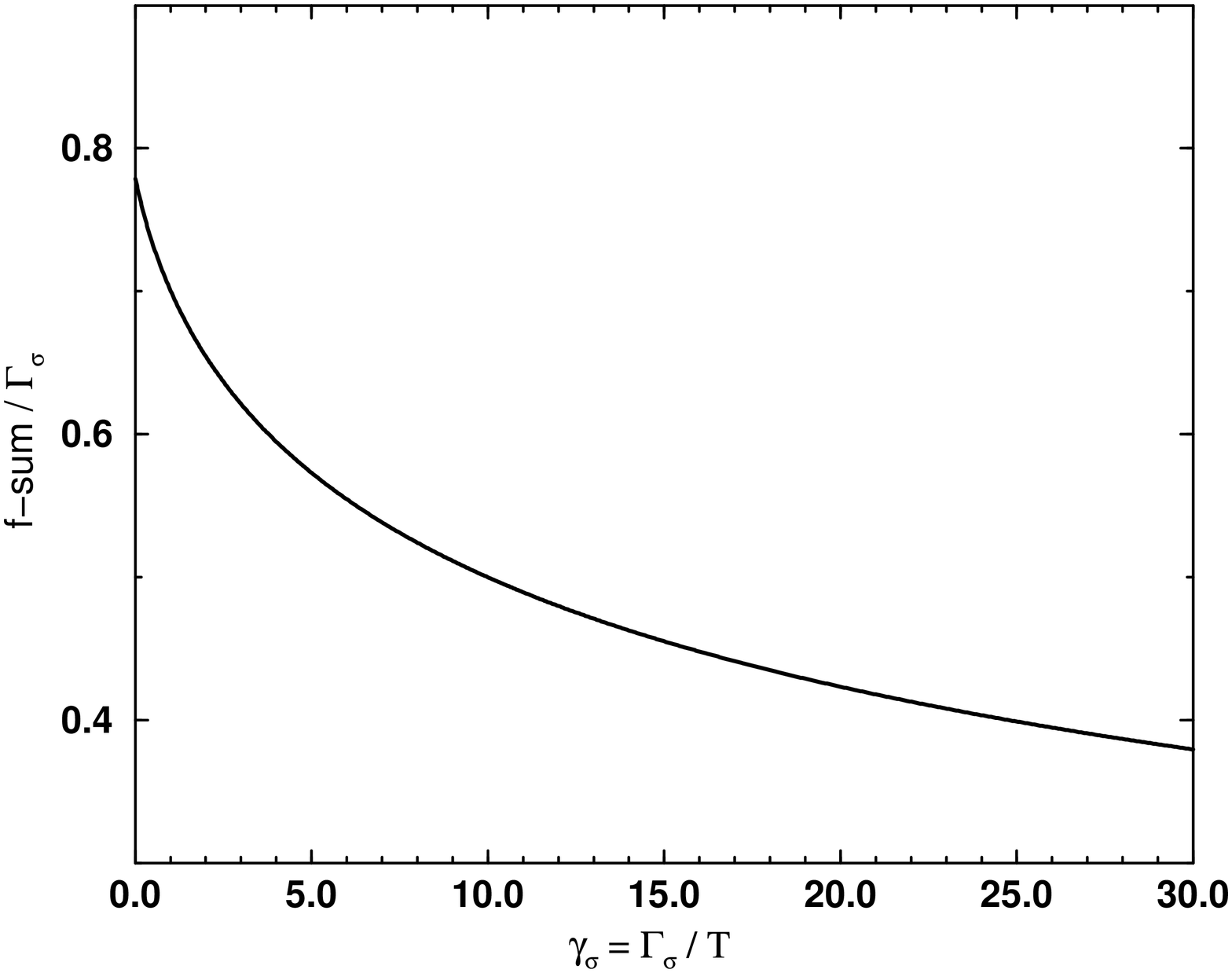,width=6.0cm,angle=270}\hfil}
\smallskip
\caption{\label{fig:fsum} $f$-sum of $S_\sigma(\omega)$ 
as defined in Eq.~(\ref{eq:fsumdefinition})
as a function of $\Gamma_\sigma$.}
\bigskip
\hbox to\hsize{\hfil\epsfig{file=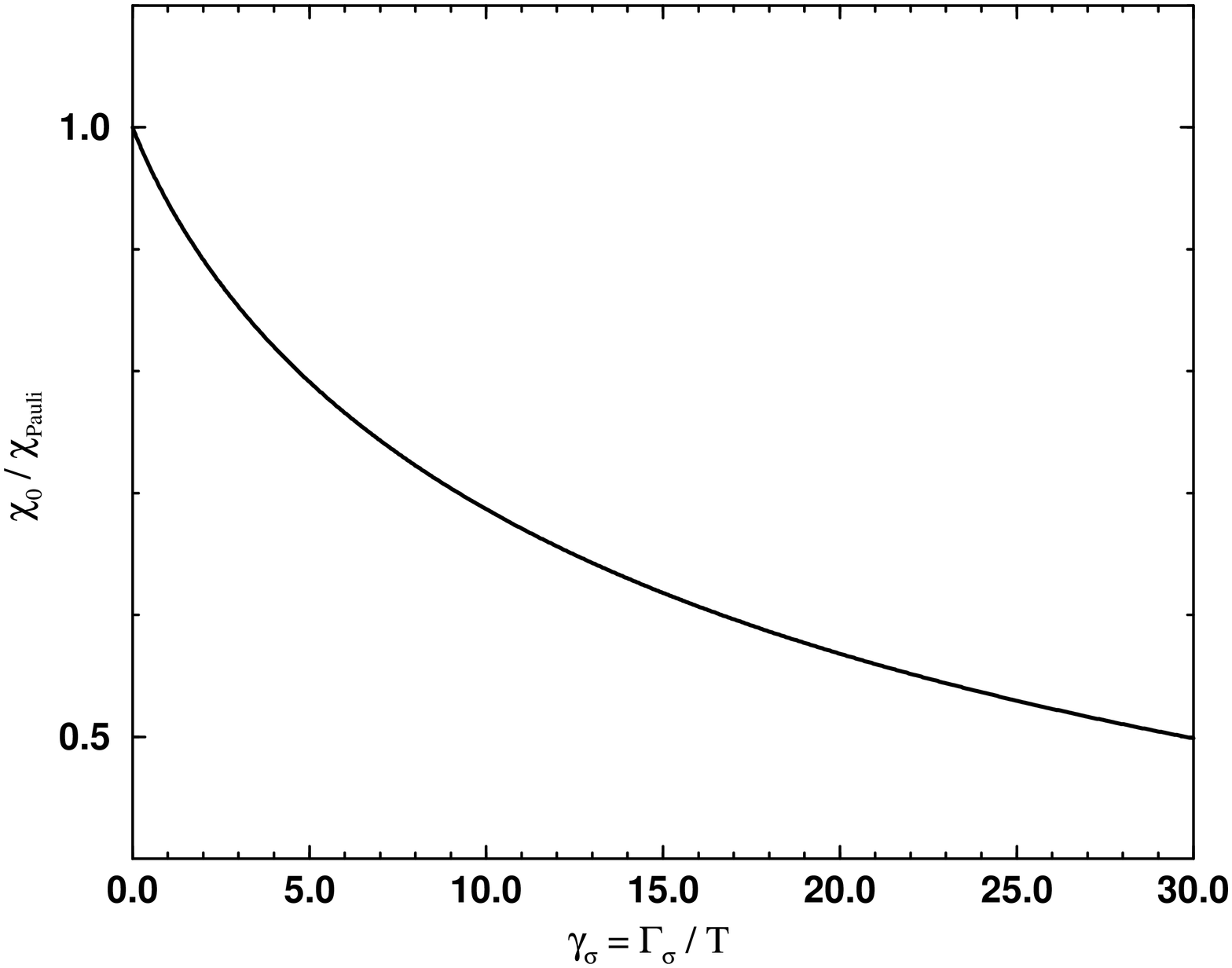,width=6.0cm,angle=270}\hfil}
\smallskip
\caption{\label{fig:static} 
Static long-wavelength limit of the magnetic susceptibility according
to Eq.~(\ref{eq:staticsus}) as a function of $\Gamma_\sigma$.}
}
\end{figure}

Our semiheuristic toy model is thus completely defined. In
Fig.~\ref{fig:norm} we show the ``downstairs $\Gamma$'' of
Eq.~(\ref{eq:resummedS}) as a function of $\Gamma_\sigma$ such that
the normalization requirement Eq.~(\ref{eq:Snorm}) is obeyed. By
construction we have $\Gamma=\Gamma_\sigma$ for $\Gamma_\sigma\to0$
with smaller values for a larger $\Gamma_\sigma$.  This reduction is
mostly due to the detailed-balance behavior which suppresses the
classical structure function for negative $\omega$.

Further we consider the $f$-sum which is for our present model
\begin{equation}\label{eq:fsumdefinition}     
\int_{-\infty}^{+\infty} \frac{d\omega}{2\pi}\,
\omega\,S_\sigma(\omega) = \Gamma_\sigma
\int_{-\infty}^{+\infty} \frac{dx}{2\pi}\, 
\frac{x}{x^2 + \gamma^2/4}\, s(x),
\end{equation}
where $\gamma\equiv\Gamma/T$. As claimed before it is equal to the
spin fluctuation rate times a factor of order unity which is shown in
Fig.~\ref{fig:fsum} as a function of $\Gamma_\sigma$.

Finally we show in Fig.~\ref{fig:static} the static long-wavelength
susceptibility in units of the Pauli susceptibility for our toy model
according to Eq.~(\ref{eq:staticsus}). It is a slowly decreasing
function of the spin fluctuation rate.

\begin{figure}[h]
\hbox to\hsize{\hfil\epsfig{file=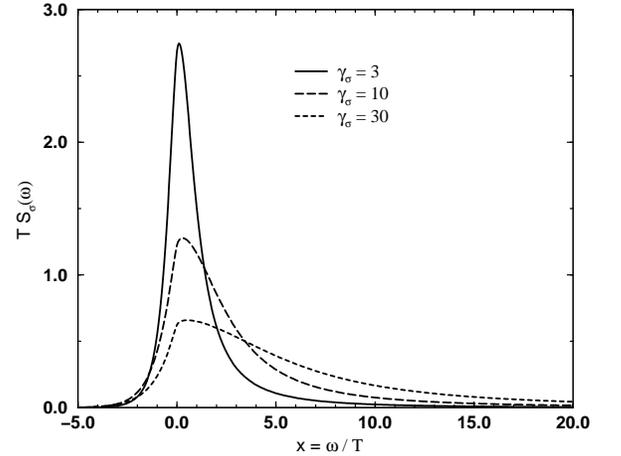,width=6.0cm,angle=270}\hfil}
\smallskip
\caption{\label{fig:heuristicS} 
Dynamical spin-density structure function $S_\sigma(\omega)$ in our
heuristic model Eq.~(\ref{eq:resummedS}) 
for $\gamma_\sigma=3$, 10, and 30.} 
\end{figure}

The overall spin-density structure function $S_\sigma(\omega)$ in our
toy model is shown in Fig.~\ref{fig:heuristicS} for several values of
$\gamma_\sigma=\Gamma_\sigma/T$.

\begin{figure}[t]
\hbox to\hsize{\hfil\epsfig{file=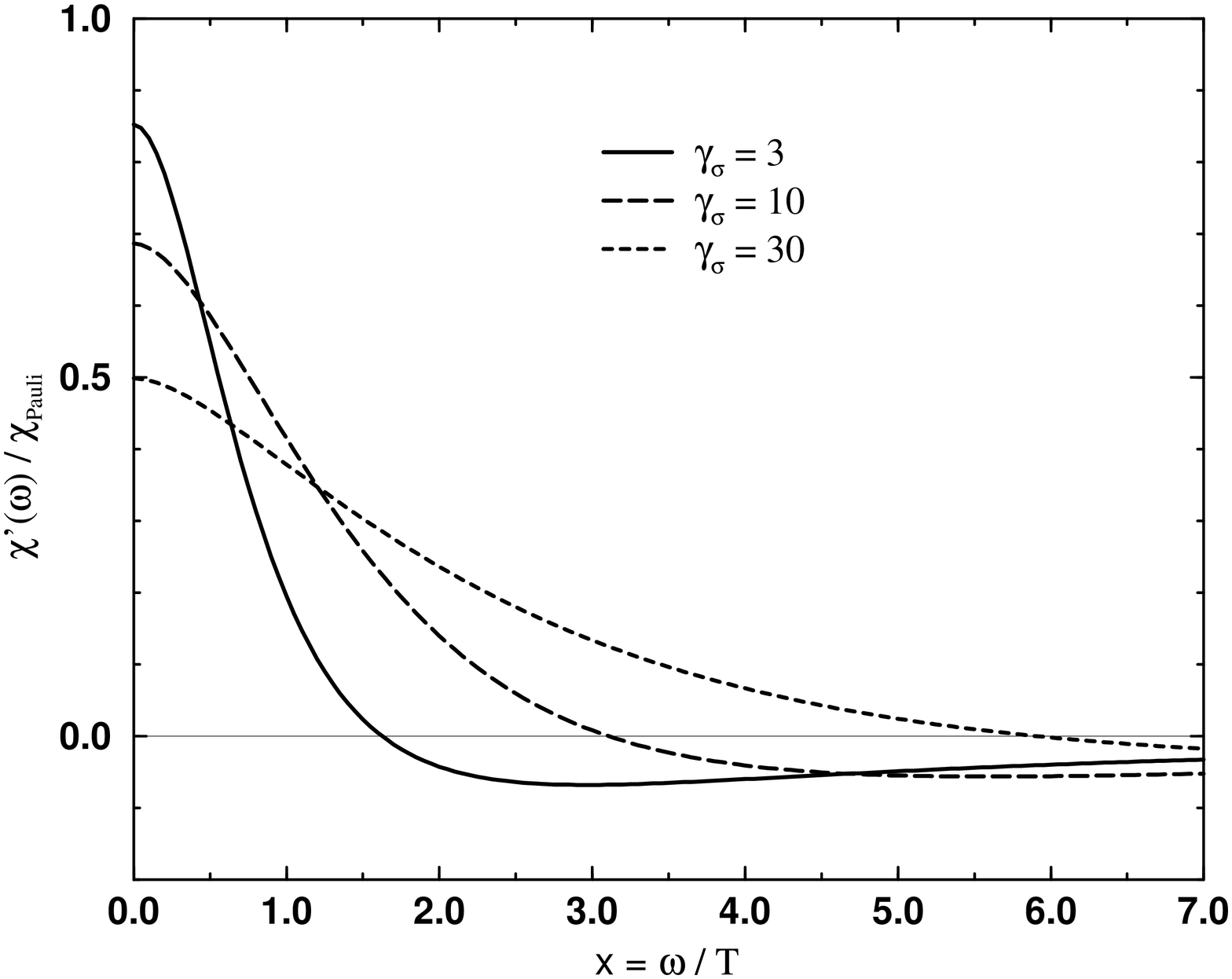,width=6.0cm,angle=270}\hfil}
\smallskip
\caption{\label{fig:chiplot} 
  Dispersive part $\chi'$
  of the magnetic susceptibility in our toy model in units of 
  $\chi_{\rm Pauli}$.}
\bigskip
\hbox to\hsize{\hfil\epsfig{file=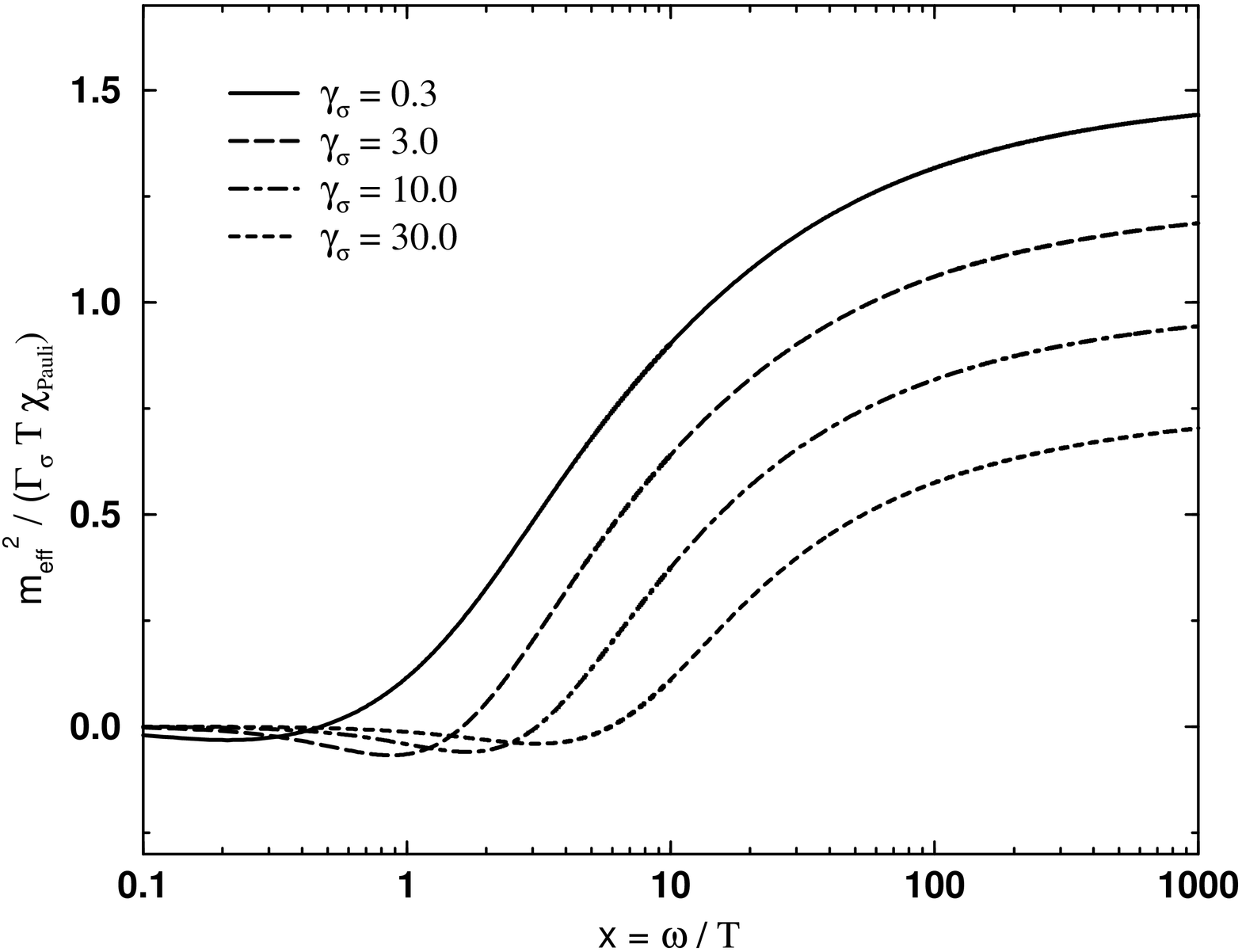,width=6.0cm,angle=270}\hfil}
\smallskip
\caption{\label{fig:meffplot}
  Effective photon mass in our toy model as a function of photon
  frequency.}
\end{figure}

Next we study the photon dispersion relation implied by our model.  In
Fig.~\ref{fig:chiplot} we show the long-wavelength limit
$\chi'(\omega)=n^2_{\rm refr}-1$ in units of the Pauli susceptibility
as a function of $x=\omega/T$ for several values of $\gamma_\sigma$.
The overall behavior is exactly as expected from our general
discussion in Sec.~\ref{sec:limitingcases}.  To see the large-$\omega$
behavior more clearly we show in Fig.~\ref{fig:meffplot} the
equivalent quantity $m_{\rm eff}^2$ in units of $\chi_{\rm Pauli}
\Gamma_\sigma T$. As predicted, $m_{\rm eff}^2$ approaches an
asymptotic value which is independent of frequency and which is of
order $\chi_{\rm Pauli} \Gamma_\sigma T$. Of course, for small
$\omega$ the squared ``effective mass'' $m_{\rm eff}^2$ begins at
negative values. However, for all frequencies and all values of 
$\gamma_\sigma$ we find that $|m_{\rm eff}^2|<m_\gamma^2$ where the
latter is the asymptotic value for $\omega\to\infty$. 

\eject

In order to arrive at a numerical estimate for the magnetically
induced photon transverse mass we write the nucleon magnetic moments
in the usual form $\mu_N=\kappa_N e/2m_N$ with $m_N$ the nucleon mass
and $\kappa_n=-1.91$ and $\kappa_p=2.79$. For simplicity we treat all
nucleons as if they were neutrons which implies 
\begin{equation}
\chi_{\rm Pauli}=4.4\times10^{-3}\,\rho_{14}/T_{10},
\end{equation} 
where $\rho_{14}\equiv\rho/10^{14}\,{\rm g\,cm^{-3}}$ and
$T_{10}\equiv T/10\,{\rm MeV}$. With the perturbative estimate
Eq.~(\ref{eq:numericalgamma}) for the spin-fluctuation rate we then
find
\begin{equation}
m_\gamma\big|_{\rm magnetic}\approx
(\chi_{\rm Pauli} T\Gamma_\sigma)^{1/2}
\approx 1.9\,{\rm MeV}\,\rho_{14} T_{10}^{1/4}.
\end{equation}
In a SN core we have densities of up to $10^{15}\,\rm g\,cm^{-3}$ and
temperatures of up to 30--$60\,\rm MeV$, implying a perturbative
spin-fluctuation rate far in excess of the temperature, i.e.\
$\gamma_\sigma=10$--100.  It has been argued that in a SN core the
true $\Gamma_\sigma$ cannot exceed a few times $T$ \cite{JKRS96,Sigl}.
Therefore, we have probably overestimated the transverse photon mass
by at least a factor of a few.

We next consider the corresponding quantity caused by the interaction
with electrons.  One finds~\cite{BS93}
$m_\gamma^2=\frac{3}{2}\omega_P^2$ with $\omega_P$ the plasma
frequency. For relativistic degenerate electrons it is
$\omega_P^2=(4\alpha/3\pi)\,p_{F,e}^2$ with $p_{F,e}$ the electron
Fermi momentum so that
\begin{equation}
m_\gamma^2\big|_{\rm electrons}=\frac{2\alpha}{\pi}\,p_{F,e}^2.
\end{equation}
Numerically, this corresponds to 
\begin{equation}
m_\gamma\big|_{\rm electrons}
=16.3~{\rm MeV}\,\,Y_e^{1/3}\,\rho_{14}^{1/3},
\end{equation}
where $Y_e$ is the number of electrons per baryon. Evidently in the
center of a SN core with $\rho_{14}\approx 8$, $T_{10}\approx4$ and
initially $Y_e\approx 0.3$ the magnetic moment contribution could be
almost as large as the electronic term.  However, because we have
probably overestimated the magnetic term by a factor of a few the
electrons still dominate.


\section{Discussion and Summary}
\label{sec:summary}

We have calculated the photon refractive index due to the interaction
with the magnetic moments of the nucleons. For simplicity we have
limited our discussion to nondegenerate neutrons. In the collisionless
limit the forward-scattering amplitude vanishes identically so that
the neutron magnetic moments alone do not cause any deviation of the
photon dispersion relation from the vacuum behavior~\cite{ON97}.
However, because of strong neutron spin interactions the collisonless
limit is far from justified in a SN core. On the basis of the
fluctuation-dissipation theorem and the Kramers-Kronig relation we
have derived a general expression for the photon refractive index in
terms of the dynamical neutron spin-density structure function
$S_\sigma(\omega,k)$. In an interacting medium it is a broad function
of $\omega$, in contrast to the collisionless limit where it is
proportional to $\delta(\omega)$.

We have found that for $\omega\ll\Gamma_\sigma$ (the neutron spin
fluctuation rate) the ``effective photon mass'' $m_{\rm eff}^2$ begins
with negative values $-\chi_{\rm Pauli}\omega^2$ in terms of the Pauli
susceptibility of the neutron ensemble. However, as shown in
Fig.~\ref{fig:meffplot} this function quickly turns around and then
grows asymptotically to a positive value $m_\gamma^2\approx \chi_{\rm
Pauli} T \Gamma_\sigma$. In absolute terms this ``transverse photon
mass'' is much larger than the maximum excursion of $m_{\rm eff}^2$ to
negative values. 

A numerical comparison for conditions relevant for a SN core reveals
that the transverse photon mass caused by the neutron magnetic moment
tends to be much smaller than that caused by the electron plasma
effect, except for extreme densities and low electron fractions where
the magnetic term may actually compete with the electronic one. A
numerically accurate comparison is not possible because the neutron
dynamical spin-density structure function is not known in any detail.
We have only performed a relatively schematic estimate which involved
many simplifying assumptions. However, it still appears safe to
conclude that the negative magnetic $m_{\rm eff}^2$ at small
frequencies cannot compete with the electronic plasma effect.  This
indicates that the total $m_{\rm eff}^2$ is always positive, i.e.\ the
photon refractive index is always less than 1 and it is reasonably
well estimated by the electronic plasma effect. This implies that the
Cherenkov processes $\nu\leftrightarrow\nu\gamma$ remain forbidden.

A more quantitative analysis than has been presented here requires a
better understanding of the dynamical nucleon spin-density structure
function, or more precisely, of the dynamical spin and isospin
susceptibilities of a hot and dense nuclear medium. We stress that the
$\omega$ dependence is crucial for the photon dispersion relation as
well as the neutrino opacities~\cite{JKRS96}, the static
susceptibilities alone which have sometimes been studied in the
literature are not enough.


\section*{Acknowledgments}

We thank E.~Braaten, P.~Elmfors and G.~Sigl for discussions or
comments on the manuscript.  This research was supported, in part, by
the European Union under contract No.\ CHRX-CT93-0120 and by the
Deutsche Forschungsgemeinschaft under grant No.\ SFB~375.


\end{document}